\documentclass[aps, prb, unsortedaddress,floatfix,nofootinbib, twocolumn, superscriptaddress,floatfix, longbibliography]{revtex4-2}

\usepackage{graphicx}
\usepackage{amsmath,amssymb,amsfonts,mathtools}
\usepackage{eucal}
\usepackage{comment}
\usepackage{physics}
\usepackage[percent]{overpic}
\usepackage{braket}
\usepackage{bm}
\usepackage[usenames,dvipsnames, pdftex]{xcolor}
\usepackage{arydshln}

\usepackage[colorlinks, breaklinks, 
            linkcolor=NavyBlue,
            citecolor=NavyBlue,
            urlcolor=NavyBlue]{hyperref}

\usepackage[all]{hypcap} 
\usepackage{enumerate}
\usepackage[normalem]{ulem}

\newcommand{\JD}[1]{{\color{Black} #1}}
\newcommand{\tth}{{t_\mathrm{th}}}

\date{\today}

\begin{document}
\title{Finite-size prethermalization at the chaos-to-integrable crossover}

\author{Johannes Dieplinger}
\affiliation{Institute of Theoretical Physics, University of Regensburg, D-93040 Germany}

\author{Soumya Bera}
\affiliation{Department of Physics, Indian Institute of Technology Bombay, Mumbai 400076, India}

\begin{abstract}
We investigate the infinite temperature dynamics of the complex Sachdev-Ye-Kitaev model~(SYK$_4$) complimented with a single particle hopping term (SYK$_2$), leading to the chaos-to-integrable crossover of the many-body eigenstates.
Due to the presence of the all-to-all connected SYK$_2$ term, a non-equilibrium prethermal state emerges for a finite time window $\tth \propto 2^{a/\lambda^{2/5}}$ that scales with the relative interaction strength $\lambda$, between the SYK terms before eventually exhibiting thermalization for all $\lambda$.
The scaling of the plateau with $\lambda$ is consistent with the many-body Fock space structure of the time-evolved wave function.
 In the integrable limit, the wavefunction in the Fock space has a stretched exponential dependence on distance. On the contrary,  in the SYK$_4$ limit, it is distributed equally over the Fock space points characterizing the ergodic phase at long times. 

\end{abstract}
\maketitle

\section{Introduction}
%
In an isolated interacting system, the process of thermalization under unitary time evolution happens via decoherence; the system acts as its bath, and at long times, the initial memory is scrambled.  
The Sachdev-Ye-Kitaev (SYK$_4$) model is an exactly solvable all-to-all connected interacting Majorana~(fermions) model, which shows thermalization with the fastest scrambling of information akin to a black hole~\cite{SY93,KitaevTalk,maldacena_review,mss_bound, Gu2020, Garcia2016, Kitaev2018, Polchinski2016, Altland2018, Bagrets2017}.
Consequently, the non-equilibrium long-time dynamics is ergodic over the many-body Fock space~(FS), and the wavefunctions are spatially structureless in FS.

On the contrary, a many-body localized~(MBL) system, such as the disordered spinless Hubbard model, is believed to evade thermalization at strong randomness at infinite temperature~~\cite{Nandkishore2015, Prelovsek-review-2017, Imbrie-review2017, AbaninBloch-Review-2018, AletReview2018}. 
The lack of thermalization  of the strongly disordered phase is  understood as `emergent' integrability described by an extensive number of local integrals of motion~\cite{Serbyn2013, Huse2014, Vosk2015, Imbrie-review2017}. 
From the FS perspective, an MBL phase would imply that the wavefunctions occupy a vanishing fraction of the FS sites~\cite{MaceMultifractality2018, Tikhonov20, tarziaFSPRB20, DetomasiPRB21, RoyFSPRB21, OritoMFPRB21}. 
However, a consensus is yet to emerge about the existence of the MBL phase in finite-size numerical studies; a slow tendency towards thermalization has been observed at strong disorder~\cite{Bera2017, Weiner19, SirkerPRL20,SirkerPRB21,MaximilianAP21, SierantPRB22, SelsBathPRB22,  Morningstar2022, Evers23}, as well as a shift in the `critical' disorder with increasing system size is identified that presumably indicates ergodicity in the thermodynamic limit~\cite{Vidmar2020, Panda2019, SierentPRL20, SierantLargeWc20, SuntasPRB20, Polkovnikov2021, AbaninAOP21}.

To address both these extreme limits, recently, a modification of the original maximally chaotic SYK$_4$ model  has been proposed by adding an integrable two-body all-to-all coupled term~(SYK$_2$). The model describes an interaction-mediated thermalizing phase and disorder-driven localization in the diagonal basis of the SYK$_2$~\cite{MonterioMBL21, GarciaPRL18}.
Even though both the terms individually are maximally entropic, the combination of the two 
allows correlations between FS site energies, essential for interpolation between the two extreme limits of Poisson~(integrable limit) and the Wigner-Dyson~(ergodic, chaotic) energy level statistics~\footnote{Recently, it is shown that in specific random graph models, the correlation between site energies is crucial to observe  a delocalization-localization transition with a critical disorder strength that scales as $\sqrt{K}$, where $K$ is the connectivity of the graph~\cite{RoyPRL20}.}. 
The model can be considered as a disordered quantum dot and is particularly suitable for controlled analytical studies due to its extensive connectivity between the FS nodes.

Additionally, \textcite{MonterioMBL21} showed that the composite model~\eqref{ham} hosts intermediate non-ergodic regimes, where the wavefunction occupies only a fraction of the entire FS.  For instance, such intermediate finite size non-ergodic regime is seen near the putative MBL transition~\cite{Luitz2016, Bera2017, Weiner19}, and in the Anderson model on random regular graphs~\cite{TikhonovRRG16, BeraRRG18, TikhonovPRB19, garciaPRR20, detomasiRRGPRB20, MirlinAOP2021}. 

A qualitative understanding can be made of the deformed SYK model~\eqref{ham} following Ref.~\cite{MonterioMBL21}. 
In the ergodic regime, the many-body energy bandwidth $\Delta_4$ of the SYK$_4$ term is larger than the bandwidth  $\Delta_2$ of the SYK$_2$ term. Therefore hybridization of states happens over the $\Delta_4$, and the states are ergodic over the entire Hilbert space.
With decreasing $\lambda$, the $\Delta_2>\Delta_4$ and less number of states hybridize, which results in a reduced Fock space where 
states resonantly coupled to a fraction of sites in the FS for a given energy.
Nonetheless, within the energy shell constructed out of these resonant sites, the state remains distributed uniformly~\cite{MonterioNEEPagePRL21}.
At even smaller $\lambda$, the number of resonant states becomes negligible, and the wavefunction localizes in the diagonal basis of the SYK$_2$ term.

Similar non-ergodic regimes have been observed in a model where the random infinite range (SYK$_4$) interaction is restricted to density-density interaction~\cite{ourwork}. Even though the interaction Hamiltonian for itself has qualitatively different properties than the original SYK$_4$ interaction, the single particle term competes to separate a fully extended phase from a localized one by regimes where wave functions are partially extended in the FS~\cite{MonterioNEEPagePRL21}. 

\begin{figure}[t]
    \centering
    \includegraphics[width=1\columnwidth]{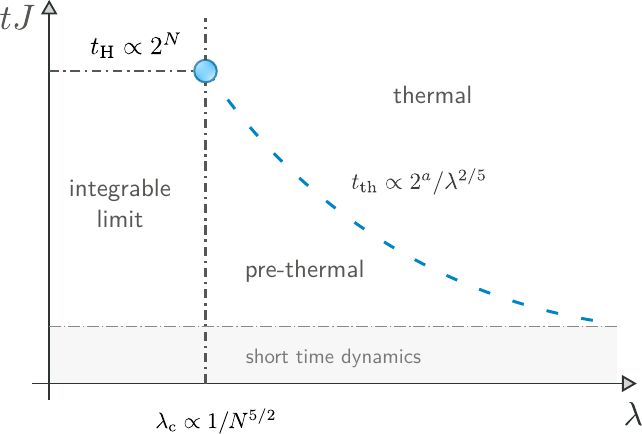}
    \caption{Qualitative dynamical phase diagram of the model Hamiltonian~\eqref{ham} as a function of the relative interaction strength $\lambda$ between the SYK terms. $N$ quantifies the number of the sites in the quantum dot. $t_\text{H}$ denotes the Heisenberg time, and $t_\text{th}$ indicated via the dashed line is the thermalizing time, which depends on $\lambda$.  }
    \label{f1}
\end{figure}

In this work, we address the following question: what is the infinite temperature dynamical properties of the crossover regime of the composite model~\eqref{ham}? 
Our observables are the density-density correlator and the wave-function propagation in the many-body FS.  
The crossover regime shows prethermal behavior at the intermediate time that scales with the relative interaction strength $\lambda$, implying thermalization in the thermodynamic limit. 
A similar conclusion can be made from the FS dynamics: a finite size crossover is observed where the critical interaction strength scales as $\lambda_\mathrm{c} \propto 1/N^{5/2}$, where $N$ is the number of sites in the quantum dot. \cite{MonterioMBL21,MirlinAOP2021}
A qualitative `dynamical' phase diagram is shown in Fig.~\ref{f1}. 

\section{Model and Method}
\paragraph*{Model.} We consider the Hamiltonian,
\begin{equation}
    \mathcal{H}=\lambda\mathcal{H}_{\text{SYK}_4}+(1-\lambda)\mathcal{H}_{\text{SYK}_2},
    \label{ham}
\end{equation}
where $\lambda\in[0,1]$ quantifies the relative strength of the two parts of the Hamiltonian and
$$
    \mathcal{H}_{\text{SYK}_4}=\sum_{ijkl}^N J_{ijkl} c^\dagger_ic^\dagger_j c_k c_l,
$$
and 
$$
    \mathcal{H}_{\text{SYK}_2}=\sum_{ij}^N t_{ij} c^\dagger_ic_j,
$$
where $c^\dagger_i$ ($c_i$) are complex fermionic creation (annihilation) at site $i\in \left[1,2,\dots,N\right]$, and $t_{ij}$ and $J_{ijkl}$ are real random numbers with zero mean and variance $\langle t_{ij}^2\rangle_\text{dis} = J^2/64N$ and $\langle J_{ijkl}^2\rangle_\text{dis} =J^2/2N^3$, with symmetry properties such as to ensure hermiticity of the resulting Hamiltonian. $\langle\, \cdot\, \rangle_{\text{dis}}$ denotes the ensemble average over independent disorder realizations of the couplings $J_{ijkl}, t_{ij}$. The scaling of the variance takes care of the proper extensive scaling of the many-body energy. 
The Hamiltonian in Eq.~\eqref{ham} has a conserved global particle number. Henceforth, we will restrict ourselves without losing generality to the filling factor $n_{\text{fill}}=1/2$.

The two parts of the Hamiltonian considered in Eq.~\eqref{ham} support a many-body chaotic and a many-body integrable phase, respectively. In the chaotic point $\lambda=1$, fully ergodic dynamics is expected, while at the integrable point, $\lambda=0$, the infinite temperature dynamics would be non-thermalizing. In the presence of both the terms, we expect a crossover between those two regimes as a function of the interaction parameter $\lambda$~\cite{GarciaPRL18, MonterioMBL21}; here, we probe the consequence of the intermediate regime in non-equilibrium dynamics.

Models of SYK$_4$+SYK$_2$ varieties have been subject to several studies in recent years in various contexts~\cite{GarciaPRL18, Haque,Lunkin,Altland_criticality,HaldarPRR20, MonterioMBL21,Dario22,LarzulPRB22}. {For instance, \textcite{Dario22} studied the adiabatic gauge potential and the spectral form factor in the modified SYK Hamiltonian. The authors found that the Thouless time scales with the system size, which is compatible with an ergodic regime for large enough system sizes at long times.
In the current work, we support the above observation with a (semi-)analytical understanding of the prethermal timescale and show -- using exact numerics -- that the real space density correlator and the FS dynamics are consistent with the predicted scaling, therefore providing a complimentary understanding of the dynamics of the mass deformed SYK model~\eqref{ham}.}

\paragraph*{Observable.} We consider the infinite temperature density-density correlator and its sample-to-sample fluctuations at finite times,
\begin{equation}
    \mathcal{C}_{i}(t)=\langle \mid \langle n_i(t) n_i(0) \rangle_\infty -n_{\text{fill}}^2\mid \rangle_\text{dis}.
    \label{eq2}
\end{equation}
Here, $n_i(t)=e^{-i\mathcal{H}t}\,c_{i}^\dagger c_i \,e^{i\mathcal{H}t}$, where the site index $i=0$ without loss of generality, since the model is fully structureless and all sites are equivalent up to disorder fluctuations, and $\langle\,\cdot\,\rangle_{T}=\Tr(e^{-\mathcal{H}/T}\,\cdot\,)/\Tr(e^{-\mathcal{H}/T})$ is the quantum mechanical expectation value at temperature $T$.

The time evolution is performed using the Kernel Polynomial method, with a Chebyshev expansion of the time evolution operator $e^{-i\mathcal{H}t}$~\cite{Wei06}. Traces are evaluated stochastically employing the principle of quantum typicality~\cite{Wei06, Bera2017}. Calculations in the limit $t\to\infty$ have been performed using exact diagonalization.

\begin{figure*}[!t]
    \centering
    \includegraphics[width=\linewidth]{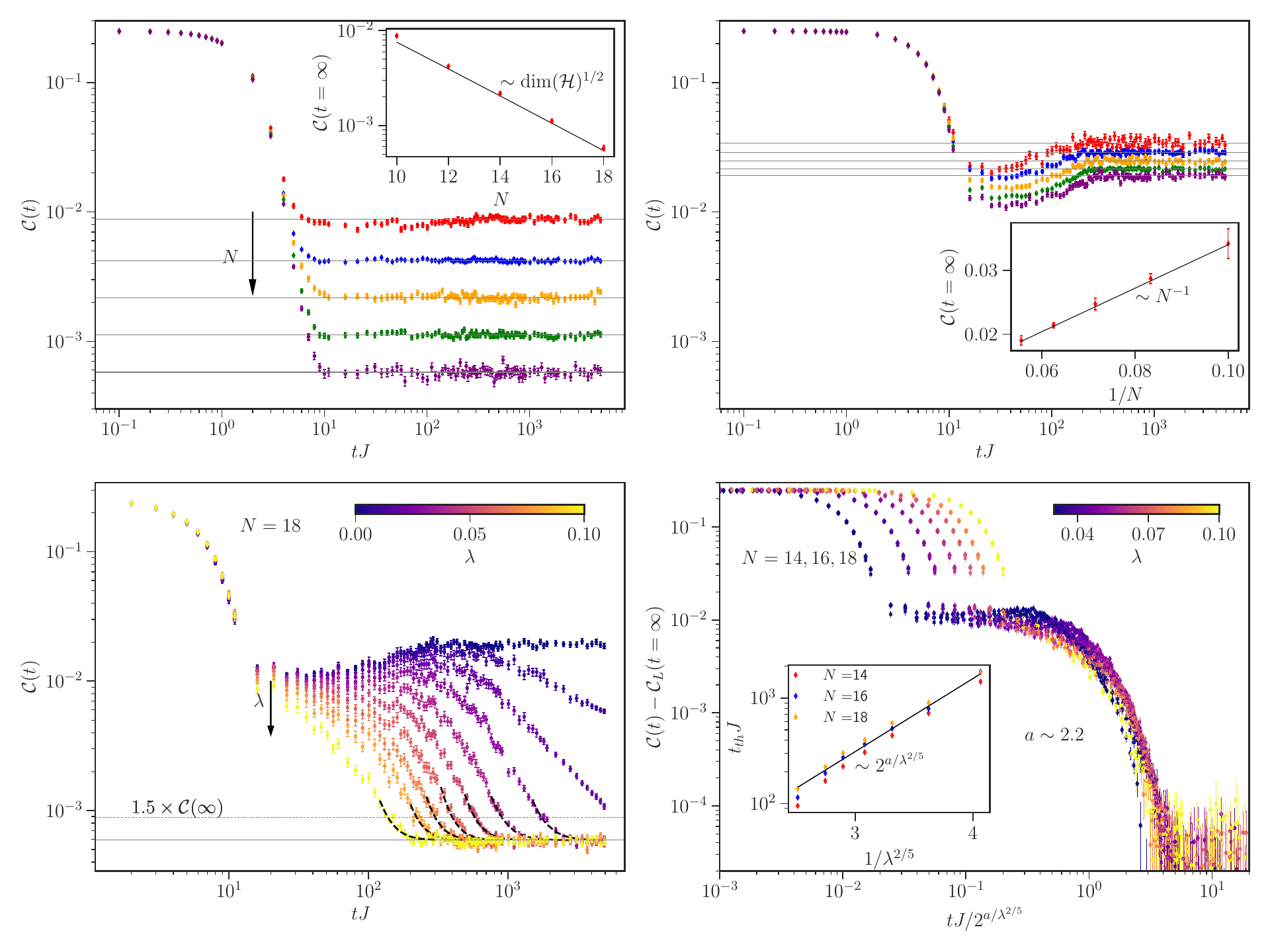}
    \caption{Density-density correlator of the SYK model with random single particle perturbations as in Eq. \eqref{ham}. (upper left) $\mathcal{C}(t)$ for the bare SYK case, i.e., $\lambda=1$, for system sizes $N=10,12,14,16,18$. The curves saturate at a plateau for $t\to\infty$, the saturation value $\mathcal{C}(t=\infty)=\langle \mathcal{C}(t>t_{\text{thresh}})\rangle_t$ is shown in the inset as a function of the system size $N$. (upper right) $\mathcal{C}(t)$ for the bare single particle case, i.e. $\lambda=0$, for the same system sizes. Also, here a plateau forms, whose saturation value is shown in the inset and scales linearly with system size $N$. (lower left) $\mathcal{C}(t)$ for different values of $\lambda=0.0,0.02,0.03,0.04,0.05,0.06,0.07,0.08,0.1$ for system size $N=18$. As a function of $\lambda$ an intermediate plateau forms, which eventually drops to the bare SYK saturation value as long as $\lambda$ is big enough. The dashed horizontal line marks the threshold value $1.5\mathcal{C}(\infty)$ below which we consider the system thermalized, i.e., we extract the thermalization time $t_{th}$ from the intersection of the fitted data (black dashed lines, for details see Appendix). (lower right) Scaling collapse of $\mathcal{C}(t)$ for different system sizes $N=14,16,18$ and $\lambda=0.03,0.04,0.05,0.06,0.07,0.08,0.1$ at long times. The inset shows the thermalization time $t_{th}$ extracted as in the previous figure, together with the estimated behavior as a function of $\lambda$ resulting from the approximate scaling collapse.
    }
    \label{f2}
\end{figure*}
%
\section{Results}
In the following, we will show the existence of a finite time window on the ergodic side of the crossover of model \eqref{ham} that is reminiscent of the slow thermalization behavior on the ergodic side of an (finite size) MBL phase. However, the thermalization process is qualitatively different; instead of a slow power-law decay e.g., the return probability, towards equilibrium~\cite{Weiner19}, we observe the forming of a finite-time plateau as we approach the integrable regime.
Additionally, we will quantify the cross-over time scale to the thermodynamic limit, in which thermalization is observed, and provide physical intuition about the FS  dynamics on both sides of the crossover.

\subsection{Density-density correlator}
\paragraph*{Limiting cases.} We begin with discussing the limiting cases in which there is either no interaction ($\lambda$=0) or no single particle term $(\lambda=1)$. 
The latter case is shown in Fig.~\ref{f2} (upper left) for different system sizes. 
In the ergodic phase, we expect the density correlator to become trivial and decoupled in the limit $t\to \infty$. 
This would correspond to $\mathcal{C}(t)\to 0$. As the systems considered are not infinite, there are still fluctuations for a given sample. Therefore, we still measure a finite $\mathcal{C}(t\to\infty)$ for a finite system size $N$ even though the model is fully ergodic. This fact, however, manifests in the scaling of the infinite time limit: As $N$ increases, the sample-to-sample fluctuations decrease with a power of the FS dimension (Fig.~\ref{f2} upper left inset). This implies that the time-evolved states explore the entire FS, and hence are fully ergodic.

In contrast to this, the non-interacting limit shown in Fig.~\ref{f2} (upper right) the correlation function  $\mathcal{C}(t)$ reaches a plateau for $t \to\infty$. However, there are two qualitative differences to the former case: On the one hand, the value of the limiting plateau is orders of magnitudes higher than in the former case.
On the other hand, the decrease of the limiting plateau with system size is qualitatively different: It follows a power law of the system size instead of the dimension of the FS (inset). It implies that even though the time-evolved states explore, in absolute numbers, more and more basis states of the FS due to its extensive connectivity, its fraction decreases with increasing system sizes exponentially. In the thermodynamic limit, the dynamics is integrable in this sense.

\paragraph*{Dynamics at the finite-size crossover.} Considering now both single particle and interaction contributions, i.e., $0<\lambda<1$, we expect a change between the two extreme cases discussed above. In Ref.~\cite{MonterioMBL21} using the statistics of the eigenstates {along with analytical calculation, the authors predicted} a finite size transition between the two regimes to happen at $\lambda_c\propto 1/N^{5/2}\ln N$. In our work, we find it difficult to resolve the $\ln N$ correction due to the unavailability of large system sizes in exact numerical simulations. Hence, in the remainder of the present article, we will compare to the leading order $\lambda_c\propto 1/N^{5/2}$.

\paragraph*{Pre-thermal plateau at intermediate times.} In Fig.~\ref{f2} (lower left) we show $\mathcal{C}(t)$ for different interactions strengths $\lambda$. For large interaction strength, the correlator follows a similar trajectory as the pure SYK model ($\lambda=1$); however, as $\lambda$ becomes smaller, a plateau develops for small enough $\lambda$ even with an additional local maximum at finite times. For those intermediate times, the data seems to resemble the non-interacting limit. One expects that one has already crossed the critical $\lambda_c$ below which we expect single particle physics, and hence integrability. However, as the system evolved to larger times, the pre-thermal plateau again dropped up to the thermal value, which has been observed in the  interacting limit (Fig. \ref{f2} upper left). This behavior suggests that even in the ergodic phase of the model for $\lambda>\lambda_c$ there is a cross-over time scale above which the time traces resemble the infinite time phase and  below which the model behaves as if it was an integrable system. This fact is the first main observation of the present work.

If $\lambda$ is decreased even more, we expect to undergo the crossover to the integrable phase, for a fixed $N$, and $\mathcal{C}(t)$ will saturate at larger values which scale with $N$ as observed in Fig.~\ref{f2} (upper right). The infinite time behavior above and below the crossover will be discussed later.

\paragraph*{Cross-over time scale.} As we have established that for the ergodic phase, there is indeed a cross-over time scale from prethermal localization to eventual thermalization, we wish to quantify this thermalization time scale $t_{th}(\lambda)$ in the vicinity of the  crossover $\lambda_c$ as a function of $\lambda$.

An analytical argument can be made from the following considerations: As we consider a finite system, the level spacing of the spectrum will provide a fixed upper bound for any physical time scales where cross-overs can happen. This usually is the Heisenberg time $t_\text{H}$ which scales with the dimension of the Hilbert space given by $t_\text{H}\approx \mathcal{O}(1)\cdot\binom{N}{N/2}$ at half filling.  Now, assuming that at the finite size transition $\lambda_c$, the crossover to the ergodic behavior happens at the latest possible time, such that below $\lambda_c$ the plateau remains present up to the $t\to\infty$ limit, we can assume  $t_{th}(\lambda=\lambda_c)=t_H$.

As for large $N$ we can approximate $t_H\propto \binom{N}{N/2}\propto 2^{b\cdot N}$, a natural assumption is to have 
\begin{equation}
    t_{th}(\lambda)\propto 2^{a/{\lambda}^{2/5}},
    \label{timescale}
\end{equation}
where $a=\mathcal{O}(1)$, which satisfies the above condition.

In Fig.~\ref{f2} (lower right), we analyze the data discussed before with respect to this analytical idea. To do so, we rescale both the time axis with the newly found $t_{th}$ and the $\mathcal{C}$-axis with the $t\to\infty$ limiting value for a given system size. This allows us to show time traces associated with different $\lambda$ and system sizes $N$ in the same figure, where -- assuming the analytical estimate is correct -- we expect them to collapse in a certain time window.

Indeed, when setting the open constant $a\approx2.2$, we observe an approximate collapse for times around $tJ/2^{a/{\lambda}^{2/5}}\approx \mathcal{O}(1)-\mathcal{O}(10)$. Additionally, we attempt to measure the time scale $t_{th}$ by defining a threshold value slightly larger than the thermal value $\mathcal{C}(\infty)$, below which we consider the system thermalized (Fig. \ref{f2} lower left, dashed horizontal line). The time scale associated with the crossing of this threshold is shown in Fig. \ref{f2} (lower right, inset), as a function of $1/{\lambda}^{2/5}$ for different system sizes. 
The theoretical prediction (black line) in Eq.~\eqref{timescale} describes the measured values reasonably well. 
\JD{The deviations from the predictions and the collapse in Fig.~\ref{f2} may be explained by the logarithmic scaling corrections of the cross-over scale of the interaction strength $\lambda$.
Additionally, the different regimes close to the finite size cross-over observed in Ref.~\cite{MonterioMBL21} may affect the thermalization properties. To resolve this a detailed study of the energy-resolved dynamics in these regimes has to be performed.}

Essentially, we have described the finite time and finite size integrable-to-chaos cross-over universally for a large range of interaction strengths largely independent from the system size (up to logarithmic corrections), with a reasonably accurate analytical estimate of the timescale $\tth$. This is the second main message of the present work.

\begin{figure}[!t]
    \centering
    \includegraphics[width=1\columnwidth]{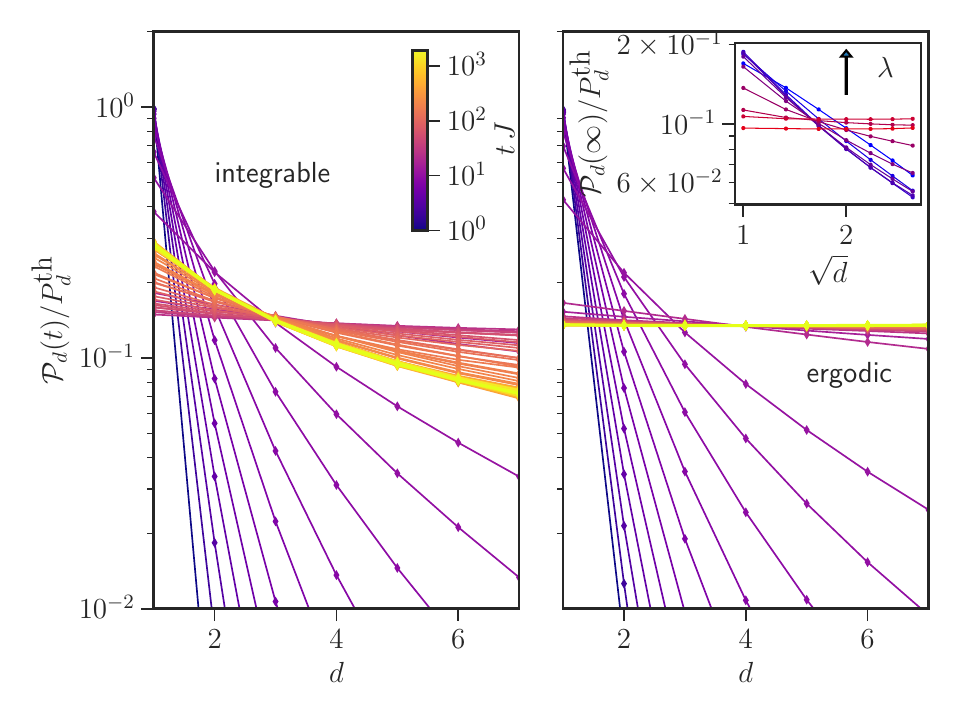}

    \caption{Time dependent probability density as a function of the FS distance to the initial state. The colors indicate the time steps (colorbar), corresponding to the time steps {shown in Fig. \ref{f2}} in the single particle case ($\lambda=0$, left) and in the fully interacting case ($\lambda=1$, right). The yellow data shows the infinite time extrapolation. The inset shows the infinite time extrapolation for different $\lambda$ (blue to red). $N=16$. 
    }
    \label{fig_fock}
\end{figure}
{The cross-over time scale between the prethermal plateau and thermalization can be understood in terms of a Thouless time, which defines the time scale after which the dynamics of the many-body system become universal, i.e., loses memory of its initial condition. We observe that the thermalization time $\tth$ approaches the Heisenberg time $t_\text{H}$ as the critical interaction strength $\lambda_c$ is approached for finite $N$.  This is similar to spin chain models with the diagonal disorder that shows an apparent lack of thermalization at strong disorder~\cite{SchiulazPRB19}. In the composite model~\eqref{ham} the role of disorder is played by the relative interaction strength $\lambda$. }

\subsection{Fock space dynamics}

In this section, we analyze the structure of the time-dependent wavefunction in the many-body FS. In the ergodic regime, we expect the wavefunction to distribute over all the FS lattice uniformly at a sufficiently long time. At the same time, in the localized phase, it should have finite support in a smaller portion of the FS lattice that does not scale with lattice size.
To this end, we start with considering the time evolution of an initial state which is fully localized on a basis state of fixed occupations at half filling.
$
    \psi(t)=e^{-it \mathcal{H}}\psi_0=e^{-it \mathcal{H}}\ket{0,0,0,\dots,1,1,1\dots}.
$
We structure the FS according to the local occupations of the basis states $\psi_a=\ket{n^a_1,n^a_2,\dots,n^a_L}$, where $n^a_i=0,1$ are the local fermionic occupations.
The (integer) distance between two basis states is defined as an FS distance.
\footnote{The distance defined here need not be the shortest path in the FS lattice. However, for our purpose, the defined measure is sufficient to measure the propagation of wavefunctions through FS.  Here it is important to note that the connectivity of the FS nodes is also scaling with the system size $N$ due to SYK interaction.}
between the string of occupations, i.e.
$
    d_H(\psi_a,\psi_b)=\frac{1}{2}\sum_{i=1}^N \abs{n_i^a-n_i^b},
$
which reaches up to a total length 
$d_H^\text{max}=N/2$.

We then define the probability for $\psi(t)$ to reach states of distance $d$ as 
\begin{equation}
    \mathcal{P}_d(t)=\sum_{\psi_a, \text{ for } d_H(\psi_a,\psi_0)=d} \abs{\braket{\psi(t)\mid \psi_a}}^2.
\end{equation}

\begin{figure}[t]
    \centering
    \includegraphics[width=1\columnwidth]{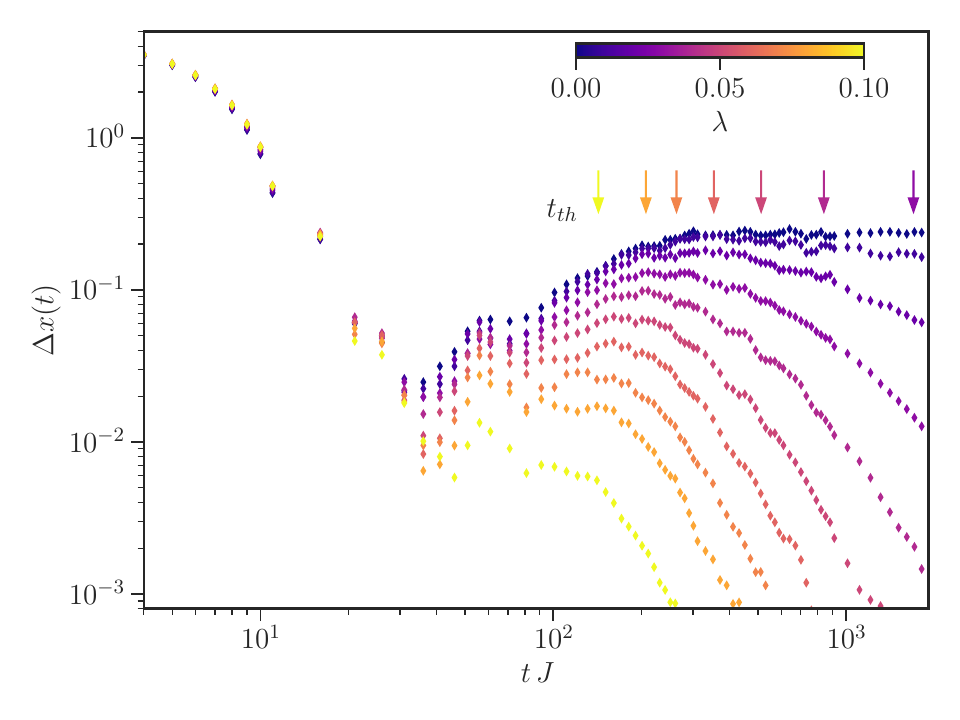}
    \caption{The first moment $\Delta x(t)$ of the distribution $\mathcal{P}_d(t)$ as a function of time for different interaction strengths $\lambda$ (color bar indicates the strength). The curves closely resemble the crossover behavior observed in the density-density correlator, Fig~\ref{f2}. \JD{The colored arrows indicate the thermalization time extracted from Fig.~\ref{f2} (inset).} The system size is $N=18$.  } 
    \label{fig_dx}
\end{figure}

For the limiting cases, i.e., single particle vs. interacting, we show the measured probabilities in Fig.~\ref{fig_fock}. For better comparability, we normalize the probability density by the equilibrium distribution,
$$
   \mathcal{P}_d^{\text{th}}=\frac{1}{\mathcal{N}_\text{th}}\binom{N-n}{d}\cdot\binom{n}{n-d} ,
$$
for filling factor $n_\text{fill}=n/N$ and normalization constant $\mathcal{N}_\text{th}=2^N\Gamma(N+1/2)/\sqrt{\pi}\Gamma(N+2/2)$ at half-filling, which corresponds to the probability density assuming equal probability on each FS, such that every distance is equally weighted.

\paragraph*{Long time structure of wave function.} Figure~\ref{fig_fock} shows the time evolution of the wavepacket in FS towards an equilibrium distribution which indeed is qualitatively different in the two limiting cases: In the integrable case ($\lambda=0$, the two body all-to-all coupled term) we observe an imbalance of the probability weight towards the initial state, suggesting a `localization' in FS with respect to the FS distance from the initial state. In the SYK limit, we observe that the distribution is flat in the long time limit, indicating that the time-evolved state has explored the full FS. This is expected and has also been observed in FS dynamics. For instance, in Ref.~\cite{CreedRoy22}, the  probability transport has been observed in a disordered quantum spin model: In the strong disorder limit, the wave function approaches an inhomogeneous infinite time limit in space, similar to what we observe in Fig. \ref{fig_fock}, and on the ergodic side it is homogeneous.

In the inset of Fig.~\ref{fig_fock} we extend this to finite $\lambda$ at $t\gg t_\text{H}$, and we observe a crossover between the imbalanced and flat distribution as a function of $\lambda$. On the localized side of the crossover, the shape of the probability density appears to be a stretched exponential with $\mathcal{P}_d(t\to\infty) \sim e^{-\sqrt{d/\xi}} \,\cdot \mathcal{P}_d^\text{th}$, where $\xi$ is the correlation length in the FS. The stretched behavior is seen all the way to the integrable point.

\begin{figure}[!t]
    \centering
\includegraphics[width=1\columnwidth]{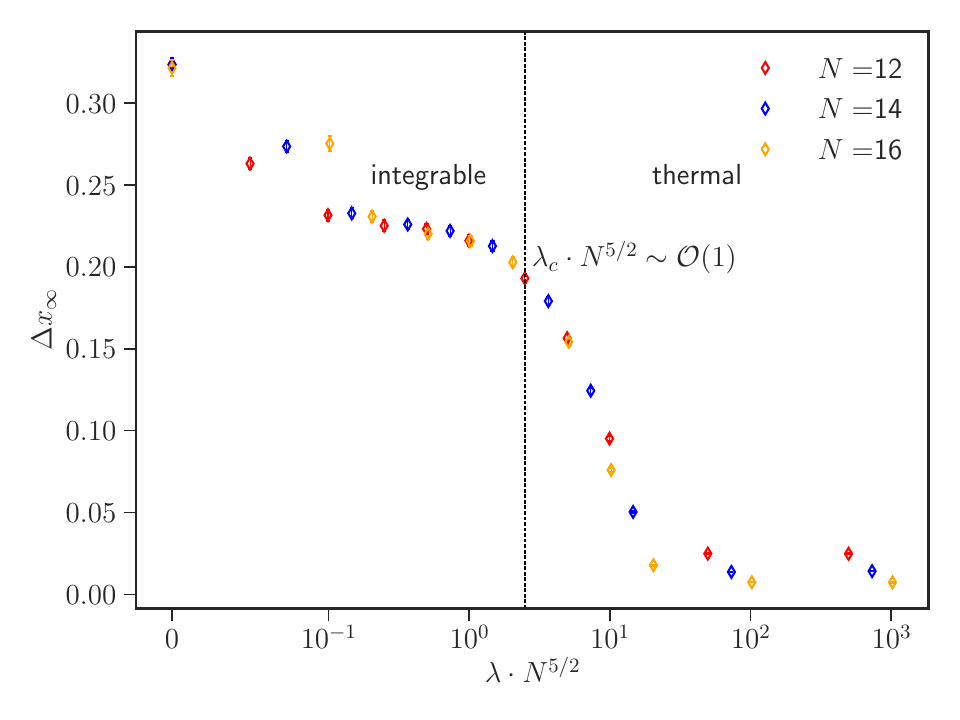}
    \caption{Finite size integrable-to-chaos crossover at infinite time. The extrapolated $\Delta x_\infty$ is a function of the rescaled interaction strength $\lambda\cdot N^{5/2}$ for different system sizes. The curves intersect approximately at the vertical dashed line (expected crossover point). 
    }
    \label{f_transition}
\end{figure}

\paragraph*{Spread of the wave function:  pre-thermal plateau.} To quantify the spread at a long time in FS, we define the first moment of the distribution with respect to the equilibrium distribution $\mathcal{P}_d^{\text{th}}$, 
\begin{align}
    \Delta x(t)&= \sum_d \left[\mathcal{P}_d(t)\,d\right] - \frac{4}{N}. 
\end{align}
With this, we characterize the extension of the time-evolved wave function in FS in the metric of the FS distance.

We show $\Delta x(t)$ for different $\lambda$ in Fig. \ref{fig_dx}. 
At short times we observe an initial decay of the form $\Delta x(t)\sim e^{-\Gamma^2 t^2}$ which is a universal feature for quantum  many-body chaotic systems~\cite{SchiulazPRB19}. The constant $\Gamma$ is related to the depletion time of the initial state. For larger times $t\gg1$, the behavior becomes non-universal, and the trajectories of $\Delta x(t)$ depend strongly on the interaction strength $\lambda$.
For large $\lambda$ the decay continues with presumably a power law, while for smaller $\lambda$, close to the crossover, we again observe the emergence of a finite time plateau, resembling the non-interacting limit, up to a thermalization time scale, where eventually in the ergodic phase thermalization occurs and $\Delta x(t)$ tends to zero. 
It is apparent that this quantity behaves qualitatively very similar to the density-density correlator in Fig.~\ref{f2}. Apparently, the FS structure imposed by the FS distance of basis states has a direct impact on the density-density correlator~\eqref{eq2}\JD{, as discussed in detail in Ref. \cite{CreedRoy22}. It has been shown that the first moment of the extension in Fock space, measured in a metric, can be related rigorously to two-point correlators in quantum spin chains.}

As a final check for this fact, we calculate $\Delta x_\infty=\Delta x(t\to\infty)$, which probes the statistics of the many-body eigenstates of the Hamiltonian which has been explored in Ref.~\cite{MonterioMBL21}. 
In Fig.~\ref{f_transition}, we observe the finite size crossover behavior as a function of $\lambda$. Indeed, there is a crossing of the data associated with different system sizes when $\lambda$ is rescaled by $N^{5/2}$. This is consistent  with the finite time dynamics and crossover we found earlier in the density-density correlator. 

\JD{The finite size scaling of the critical interaction strength $\lambda_c$ is consistent with the results found in similar models in Ref.~\cite{MonterioMBL21,MirlinAOP2021}. Additionally, it matches the prediction with random regular graphs (RRG) transition. Recently, \textcite{Mirlin2023} predicted a critical disorder $W_c^\text{RRG}\propto N^4 \log{N}$, which is related to the critical interaction strength $\lambda_c$ by $W_c\propto \lambda_c^{-1}N^{3/2}$, where the factor $N^{3/2}$ is accounted for by the different normalization of the interaction term in our model~\eqref{ham}. Quantitatively, the numerical values of the cross-over scale for a fixed system size are consistent with each other. }

\section{Discussion and Conclusion}
To summarize, our extensive numerical investigation suggests a rich dynamical phase diagram~(as indicated in Fig.~\ref{f1}) for the SYK$_4$+SYK$_2$ model~\eqref{ham} along with the finite size chaos to integrable crossover.  Our observables are the density-density correlation function and the wavefunction spread in the many-body FS~\footnote{
We note that both these quantities are related closely to the spectral form factor, which is a generic feature of 2-point correlators of SYK-related models~\cite{Altland21}. Therefore, a similar dependence of the plateau value with relative interaction strength should be expected in the spectral form factor also~\cite{Dario22}.}. 
As expected, the SYK$_4$ shows ergodic dynamics in real space and many-body FS. In the presence of the SYK$_2$ term, in the intermediate time, the model shows a pre-thermal plateau that scales with the interaction strength $2^{a/\lambda^{5/2}}$ with $a{\approx}2.2$. After the prethermal plateau for all values of $\lambda$, we observe a $\lambda$ dependent power-law decay of the return probability both in real space and in the FS.

The long-time asymptotic behavior of the wavepacket in the FS indicates a crossover at $\lambda_\mathrm{c}\propto 1/N^{5/2}$, which is consistent with the prediction of Ref.~\cite{MonterioMBL21}. In particular, we observe that in the integrable limit, the wavefunction in the FS decays in a stretched exponential ${\propto}e^{-\sqrt{d/\xi}}$ with FS distance $d$. These observations imply that for finite $N$ the SYK$_2$ term is a relevant perturbation to the SYK$_4$ physics. 

In the context of the dynamics of the finite size chaos-integrable crossover, it would be relevant to study how the states, contributing to the prethermal plateau relax into the energy shells found in Ref.~\cite{MonterioMBL21}. Here, it would be helpful to study the energy-resolved correlators as studied in spin chains~\cite{Bera2017}. We leave this for future studies.

Given the current understanding of the non-equilibrium dynamics, it is pertinent to ask whether it represents the slow thermalization physics in the context of MBL transition that we know in finite size and finite time numerical studies~\cite{Bera2015, Luitz:2017cp, Weiner19, DoggenRevAnnPhy21, Evers23}. For short-range models, the density-density correlator shows a slow propagation of charge-density even at large disorder~\cite{Weiner19}, without a hint of prethermalization at intermediate times. 
In the FS, several different transport statistics have revealed a power-law decay of return probability at intermediate disorder strengths~\cite{TorresFSPRB15, SchiulazPRB19, CreedRoy22}. Therefore, in this class of models, there is a lack of evidence for finite size prethermalize plateau.

In contrast, for models with finite range $1/r^\alpha$ interaction or hopping~\cite{BurinPRB15,GutmanPRB16, TikhonovPRB18, NagPRB19, DetomasiPRB19, RoyScipost19, ThomsomPRR20, DengPRL20} a finite-size transition has been observed similar to the composite model~\eqref{ham}. It is generically found that the critical disorder strength scales with the system size as $W_{\text{c}} \propto L^{2d-\alpha} \ln L$ for $d{<}\alpha{<}2d$ where $d$ is the system dimension. The implication of the finite range interaction~\cite{ourwork} and hopping on the observed prethermal plateau would be a relevant question for further study.

We end with the following observation: Our numerical calculation is performed on the computational basis, where both SYK$_4$, and the SYK$_2$ terms are non-diagonal; therefore, we do not expect to observe localization. It would be relevant now to study the model~\eqref{ham} in the diagonal basis of SYK$_2$, where an ergodic-to-localization transition is possible. The model can be understood as SYK$_4$ complimented with a correlated chemical potential disorder, and the dynamical properties of such a model would be generically interesting to study in the context of the existence of the MBL phase. 


\section*{Acknowledgements}
We want to thank A. Altland,  G. De Tomasi, A. Haldar, D. Joshi and A. D. Mirlin for critical reading of the manuscript and several valuable comments. We are grateful to F. Evers, and A. D. Mirlin for several fruitful discussions on topics related to localization transition. \JD{We are particularly grateful to Sasha Mirlin for pointing out an extra normalization factor in the interaction term, which was missed in a previous version of the manuscript.} We also thank D. Rosa for bringing Ref.~\cite{Dario22} to our attention.
JD acknowledges support from the German Academic Scholarship Foundation and the German Research Foundation (DFG) through the Collaborative Research Center SFB 1277.
SB acknowledges support from SERB-DST, India, through Matrics (No. MTR/2019/000566), and MPG for funding through the Max Planck Partner Group at IITB.

\bibliography{ref}

\end{document}